\documentclass[10pt, english, aps, prl, amsmath, amssymb, twocolumn,showpacs]{revtex4-1}
\usepackage[T1]{fontenc}
\usepackage[latin9]{inputenc}
\usepackage{amstext}
\usepackage{graphicx}
\usepackage{float}
\usepackage{datetime}
\usepackage{ulem}
\usepackage{color}
\usepackage{bm}

\newcommand{\LB}  {L_\mathrm{B}}

\makeatletter
\@ifundefined{textcolor}{}
{%
 \definecolor{BLACK}{gray}{0}
 \definecolor{WHITE}{gray}{1}
 \definecolor{RED}{rgb}{1,0,0}
 \definecolor{GREEN}{rgb}{0,1,0}
 \definecolor{BLUE}{rgb}{0,0,1}
 \definecolor{CYAN}{cmyk}{1,0,0,0}
 \definecolor{MAGENTA}{cmyk}{0,1,0,0}
 \definecolor{YELLOW}{cmyk}{0,0,1,0}
}

\makeatother

\usepackage{babel}
\begin{document}

\title{Finite-Size Scaling of the Density of States in Photonic Band Gap Crystals}

\author{Shakeeb Bin Hasan}
\author{Allard P. Mosk}
\altaffiliation[Current address:\ ]{Nanophotonics, Debye Institute for Nanomaterials Science, Utrecht University, The Netherlands}
\author{Willem L. Vos}
\author{Ad Lagendijk}

\affiliation{Complex Photonic Systems (COPS), 
MESA+ Institute for Nanotechnology,
University of Twente, 
P.O. Box 217, 7500 AE Enschede, The Netherlands} 

\date{Day: \today,    \ Time:\ \currenttime} 

\begin{abstract}
The famous vanishing of the density of states (DOS) in a band gap, be it photonic or electronic, pertains to the infinite-crystal limit. 
In contrast, all experiments and device applications refer to finite crystals, which raises the question: Upon increasing the linear size $L$ of a crystal, how fast does the DOS approach the infinite-crystal limit? 
We present a theory for finite crystals that includes Bloch-mode broadening due to the presence of crystal boundaries. 
Our results demonstrate that the DOS for frequencies inside a band gap has a $1/L$ scale dependence for crystals in one, two and three dimensions. 

\end{abstract}

\maketitle 
The discovery brought about by crystallography that a crystal consists of an infinite array of unit cells with periodic symmetry~\citep{Bragg1913PRS} has led to the birth of modern condensed matter physics~\cite{Ashcroft1976Book}.
The quantum-mechanical description of electronic degrees of freedom of the solid state has led to the notion of the density of states (DOS), and to the characterization of semiconductors by a frequency range of a vanishing DOS, a band gap~\cite{Ashcroft1976Book,Sheng2006Book,Akkermans2011Book}.
An analogy can be drawn between electronic condensed matter and nanophotonic condensed matter phenomena, as the underlying mechanism for the formation of a band gap in both cases is wave interference~\cite{Sheng2006Book,Akkermans2011Book,Joannopoulos2008Book}. 
Indeed, photonic crystals exhibit Bragg reflections for light~\cite{Vukusic2003N}.
When the light-matter interaction is sufficiently strong, photonic crystals develop a complete 3D band gap in the photonic DOS that is the nanophotonic analogue of electronic semiconductors and insulators~\cite{Yablonovitch1987PRL,John1987PRL,Joannopoulos2008Book}.

Most theories of the DOS in condensed matter and in nanophotonics consider infinite systems ($L\rightarrow\infty$). 
Examples are the plane-wave expansion for waves, both electronic~\cite{Ashcroft1976Book} and photonic~\cite{Joannopoulos2008Book}, or the thermodynamic limit in liquid state theory~\cite{Hansen2013Book} that all exploit the underlying periodic or continuous symmetry. 
Analytic theories for the DOS of \textit{finite-size} crystals are much more difficult to devise~\cite{Yndurain1974SSC}. 
In contrast, for many other observables, with the electric conductance as prime example~\cite{Anderson1958PR,Abrahams1979PRL,Fisher1985PRB,Sheng2006Book}, scaling as a function of system size has come to play a central role in condensed matter physics.

The concept of a band gap, electronic~\cite{Ashcroft1976Book} or photonic~\cite{Bykov1972JETP,Yablonovitch1987PRL,John1987PRL,Vats2002PRA}, applies to infinite systems only.
In contrast, experiments and device applications obviously involve finite crystals~\cite{Ogawa2004S,Leistikow2011PRL}, which raises the question: How fast does the DOS in the band gap of a finite crystal approach the infinite-crystal limit? 
To our knowledge, there is no theory that addresses a finite photonic band gap \textquotedbl{}crystal\textquotedbl{} embedded in infinite free space. 
Developing a model for such systems implies the inclusion of interfaces. 
Having such a theory is a prerequisite to assess applications of photonic band gap crystals that rely on the total DOS. 
For a photonic medium, the local density of states (LDOS) represented by  ${\rho(\omega,{\bf r})}$ is defined as the number of states per frequency per volume at position ${\bf r}$.
Integrating the LDOS over a certain volume ${\cal V}$ results in the number of states in that volume:
%
\begin{eqnarray}
{\cal N}(\omega) = \int_{\cal V} 
\rho\left(\omega,{\bf r}\right) {\rm d} {\bf r} . 
\end{eqnarray}
%
The DOS in this volume is defined as 
\begin{eqnarray}
\rho (\omega) \equiv \frac{\cal N (\omega)}{\cal V}, 
\end{eqnarray}
and the DOS of vacuum is $\rho_{\rm vac}(\omega) = \omega^2/(\pi^2 c^3)$.

Let us consider a $D$-dimensional photonic crystal of linear size ${\cal L}$ with lossless boundary conditions such as periodic boundary conditions (PBC). 
In this crystal, Bloch modes appear with a wave vector ${\bf k}$ in the first Brillouin zone, a band index $n$, and eigenfrequencies $\omega_n({\bf k})$. 
For a photonic crystal with PBC, the DOS in ${\cal L}^D$ can be calculated from the Green's function $G(\omega^2)$~\cite{Eco90, Sprik1996EPL,Busch1998PRE}:
%
\begin{eqnarray}
\rho_{\rm PBC}(\omega)
&=& 
\frac{2 \omega}{ {\cal L}^D} \left( \frac{-1}{\pi} \right) {\rm Im}  \left\{ {\rm Tr} \, G(\omega^2) \right\} \\
&=&
-\frac{2 \omega}{\pi {\cal L}^D} {\rm Im} 
\left ( \lim_{\epsilon \rightarrow 0^+}
\sum_{{\bf k},n}
 \frac{1}{\omega^2_n({\bf k} ) -\omega^2 + i \epsilon} \right),\label{eq:pbc_dos_definition1}\\
&=&
\frac{2 \omega}{{\cal L}^D} \sum_{{\bf k} ,n} \delta \left(\omega^2_n({\bf k} ) -\omega^2\right).\label{eq:pbc_dos_definition2}
\end{eqnarray}
%
As long as ${\cal L}$ is finite, $\bf{k}$ is discrete and $\rho(\omega)$ consists of a sum of delta functions separated in frequency. 
Only in the limit ${\cal L}\rightarrow \infty$ does the DOS become a \textit{continuous} function of frequency.
Apparently, crystals with finite size ${\cal L}$ and PBC do not constitute a good model for open systems. Even if we supply the finite-size crystals with boundary conditions different from PBC but still energy preserving, we will always end up with a sum of \textit{discrete} delta functions for finite ${\cal L}$.

Experiments are done on finite crystals with open boundaries or in technical terms, on crystals with finite support, with a volume ${ V_{\rm s}} = (L_{\rm s})^D $.
These crystals have a \textit{continuous} density of states for any value of the support $L_{\rm s}$ in agreement with experiments. 

When Bloch waves reach an interface, they can escape into free space, which is an absorption event. Therefore, to describe finite-support systems, we need absorptive boundary conditions.
We introduce an effective intensity reflection coefficient ${\cal R}$ that is an average of the angle-dependent reflection coefficients of Bloch modes. The magnitude of ${\cal R}$ depends on the refractive index of the surrounding free space. Basically, ${\cal R}$ describes the mixed character of the boundary conditions~\cite{Morse1953Book,Lagendijk1989PLA,Zhu1991PRA}.
The propagation length of a Bloch mode becomes
%
\begin{eqnarray}
L = \frac{1+{\cal R}}{1-{\cal R}}L_{\rm s}, \label{mean-free-path}
\end {eqnarray}
%
where $L$ is the effective linear size of the $D$-dimensional crystal.
For ${\cal R}<1$, the propagation lengths $L$ of Bloch modes are finite, and the wave vectors become complex valued:
%
\begin{eqnarray}
{\bf k} \rightarrow {\bf k} + i \hat{{\bf k}}/L, \label{eq:ksmearing}
\end{eqnarray}
%
where $L$ functions as a mean free path. From here on, we take without loss of generality ${\cal R} = 0$, implying $L = L_{\rm s}$. 
Other values of ${\cal R} $ lead to the same scaling laws with $L$.

To calculate the finite-$L$ correction to the density of states, we insert Eq.~(\ref{eq:ksmearing}) in to dispersion relation
%
\begin{equation}
\omega_n({\bf k} + i \hat{{\bf k}}/L)
\approx \omega_n({\bf k}) +  i \Delta_{n}({\bf k}), \label{eq:complexdisp}
\end{equation}
%
with 
%
\begin{equation}
 \Delta_{n}({\bf k}) \equiv 
 \frac{1}{L} \left|\hat{{\bf k}} \cdot \frac{\partial \omega_n({\bf k})}{\partial {\bf k}}\right|,
\end{equation}
%
where the absolute value ensures that the imaginary part reflects absorption and not gain. The change of the real part of the eigenfrequencies due to scattering or absorption is a higher-order correction to be included in future work.

Incorporating dispersion [Eq.~(\ref{eq:complexdisp})] into Eq.~(\ref{eq:pbc_dos_definition1}) results in
%
\begin{eqnarray}
\rho(\omega)& = &  \sum_{{\bf k},n}  \frac{2 \omega }{-\pi}
{\rm Im} \frac{1}{\omega_{n}^2({\bf k}) - \omega^2
 + i 2 \omega_{n}({\bf k}) \Delta_{n}({\bf k})  } \label{eq:pc_dos_definition_finite}  \label{eq:full_lorentzian_dos}.
 \end{eqnarray}
%
The modification caused by our model is that the sum of delta functions becomes a sum of Lorentzian line shapes. Equation~(\ref{eq:full_lorentzian_dos}) is our central result.
Remarkably, it allows us to calculate the DOS in a finite-volume photonic crystal from the band structure of an infinite crystal. 
It offers a direct way to predict volume-averaged rates for processes that depend on the DOS, such as spontaneous emission from many sources distributed over the photonic crystal.

\begin{figure}
\begin{centering}
\includegraphics[width=1 \columnwidth]{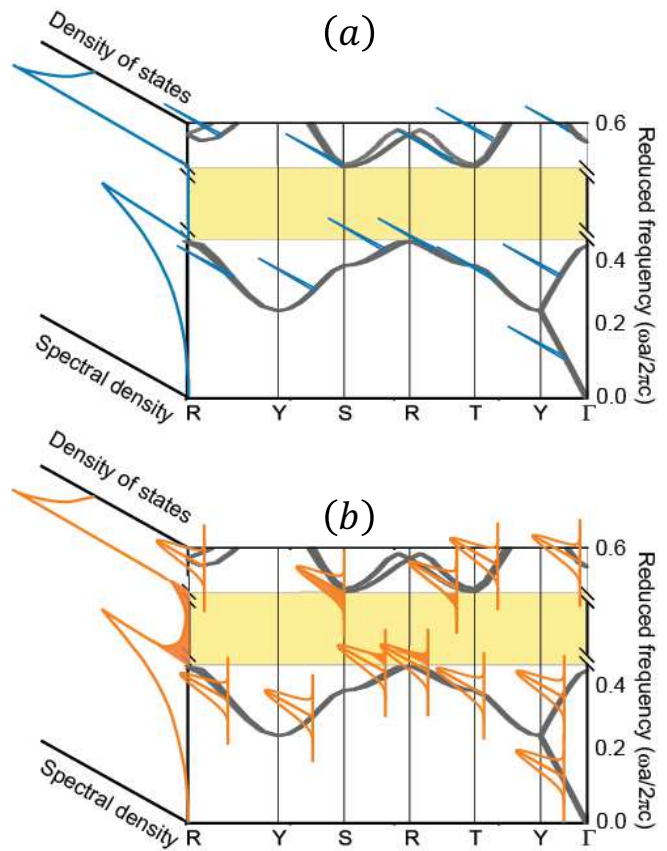}
\par\end{centering}
\caption{Schematic presentation of crystal modes that contribute to the DOS.
(a) In an infinite photonic crystal, the continuum of modes outside the band gap are represented by delta functions (blue peaks). 
The resulting DOS, plotted in the third dimension at the left, vanishes in the band gap and is only nonzero outside the gap.
(b) In a finite-support crystal the modes become Lorentzian (orange). 
Since the modes have finite widths, they extend into the band gap, thereby causing a nonzero DOS in the gap (orange-filled area). \label{fig:dos_cartoon}}
\end{figure}

Figure~\ref{fig:dos_cartoon} illustrates our model. 
In Fig.~\ref{fig:dos_cartoon}(a), the modes for an infinite crystal are depicted as zero-linewidth peaks that do not spill into the band gap. 
For the finite-support crystals, Fig.~\ref{fig:dos_cartoon}(b), the modes become Lorentzians with finite widths. 
As a result, the modes extend into the band gap, thereby causing a nonvanishing DOS.
%
%

\begin{figure}
\begin{centering}
\includegraphics[width=0.9\columnwidth]{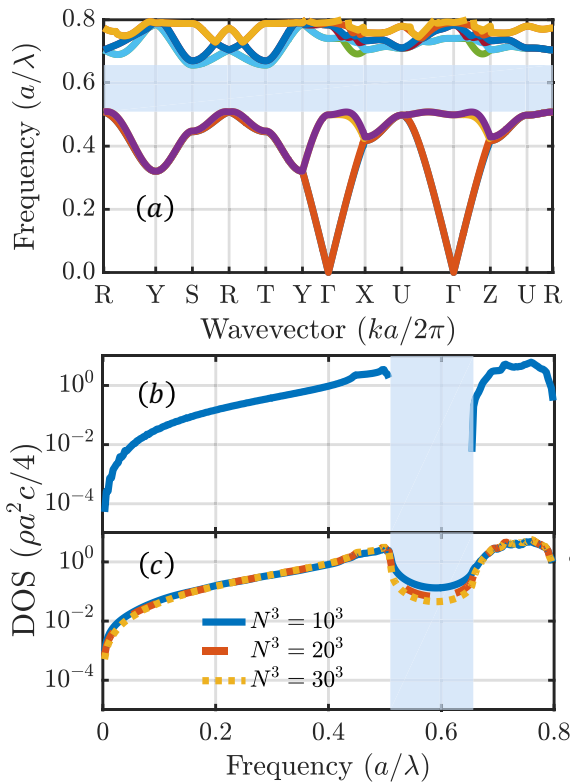}
\par\end{centering}
\caption{(a) Photonic band structure of a 3D inverse woodpile crystal. 
The lattice constant of one side of the tetragonal unit cell is $a$, while the other two are $a/\sqrt{2}$. 
Air holes in the silicon host (dielectric constant $\varepsilon = 12.1$) have a radius of $R = 0.24a$. 
The highlighted blue area indicates the band gap.
(b) DOS for the infinite crystal as a function of the reduced frequency $\tilde{\omega} \equiv \omega a / 2 \pi c =  a/\lambda$. 
The DOS is scaled with $4/(a^2c)$, leading to $\tilde{\rho}_{\rm vac} = \tilde{\omega}^2$. 
(c) DOS calculated according to Eq.~(\ref{eq:full_lorentzian_dos}) for a finite-support crystal of volume $L^3/2 \equiv a^{3}N^3/2$, where $N^3$ is the number of unit cells~\cite{SI}.}
\label{fig:DOSIwp}
\end{figure}

As a generic example of a 3D photonic band gap crystal, we consider a cubic diamondlike 3D inverse woodpile crystal that is known to exhibit a broad 3D photonic band gap~\citep{Ho1994SSC,Hillebrand2003JAP,Huisman2011PRB}. 
Figure~\ref{fig:DOSIwp}(a) shows the photonic band structure calculated using the plane wave expansion method~\citep{Johnson2001OE} in the irreducible Brillouin zone. The crystal is made of silicon (dielectric constant $\varepsilon = 12.1$) with cylindrical air holes of radius $R = 0.24a$, where $a$ is the length of one side of the tetragonal unit cell while the other two lattice parameters are $a/\sqrt{2}$~\cite{SI}. 
The DOS of the infinite crystal in Fig.~\ref{fig:DOSIwp}(b) shows the band gap. 
The corresponding DOS for a finite-support crystal with volume  $N^{3}a^{3}/2 \equiv L^3/2$, with $N^3$ being the number of unit cells, is shown in Fig.~\ref{fig:DOSIwp}(c). 
We find that the DOS inside the band gap does not vanish anymore due to the nonzero linewidth of the modes. 
Expectedly, the DOS in the band gap decreases with increasing crystal size. 

\begin{figure}
	\begin{centering}
		\includegraphics[width=0.9\columnwidth]{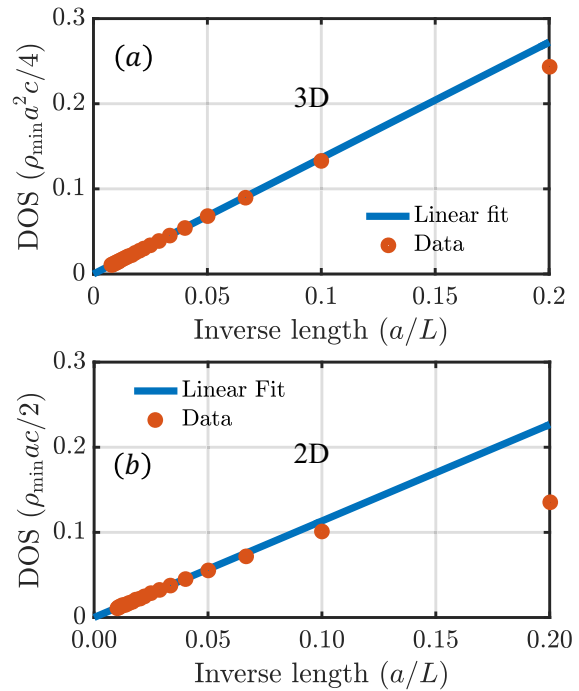}
		\par\end{centering}
	\caption{Minimum of the DOS in the band gap of (a) a 3D inverse woodpile crystal and (b) a 2D square lattice crystal (b) versus the inverse length $1/L$ of crystals.
Red circles denote the DOS calculated using Eq.~(\ref{eq:pc_dos_definition_finite}), while the solid blue curve is the linear fit in the large size limit with slopes (a) $m = 1.36$ and (b) $m=1.13$ (b).} 
	\label{fig:minDOS_IWP}
    \end{figure}

To obtain the finite-size scaling of the DOS inside the band gap, we plot in Fig.~\ref{fig:minDOS_IWP}(a) the minimum DOS in the band gap as a function of the inverse of the linear size $L$. 
Considering the exponential decrease of the LDOS at the center of a finite-size crystal as a function of its size~\cite{Leistikow2011PRL,Yeganegi2014PRB,Asatryan2001PRE,Ishizaki2009}, it is remarkable that the minimum DOS in Fig.~\ref{fig:minDOS_IWP}(a) is very accurately described by a $1/L$ dependence.
%

To investigate the dependence on dimensionality, we consider a generic 2D square lattice crystal of lattice constant $a$ made of dielectric cylinders that exhibits a band gap for the $s$ polarization (electric field out of the plane)~\cite{Joannopoulos2008Book}. 
The minimum DOS inside the band gap is shown in Fig.~\ref{fig:minDOS_IWP}(b) for a finite-size crystal of area $a^2N^2\equiv L^2$ versus inverse length $1/L$. 
As in the case of a 3D crystal, we find the DOS predicted by Eq.~(\ref{eq:full_lorentzian_dos}) to be well described by a $1/L$ scaling in a 2D crystal~\cite{SI}. It seems that deviations from the $1/L$ scaling for very small crystal sizes happen earlier for 2D than for 3D. This could be a dimensionality effect, as the surface-to-volume ratio is larger in 2D than in 3D, but could also be due to the fact that the crystal types are different.

The finite-size scaling of the DOS in 1D can be calculated exactly; hence, we do not have to use our model Eq.~(\ref{eq:full_lorentzian_dos}). 
In the Supplemental Material, we include the case of a generic 1D Bragg stack in which the DOS is calculated rigorously from the position-dependent LDOS. 
The DOS in the band gap of the 1D stack also exhibits the inverse linear scaling versus crystal size, establishing the universality of the DOS finite-size scaling inside the band gap for crystals of any dimension. 
Moreover, we have calculated the finite-size scaling for the same 1D crystal using our model of Bloch mode broadening. To our satisfaction, the Lorentz model also shows a $1/L$ dependence of the DOS~\cite{SI}. 

Exact calculations for 2D and 3D photonic crystals are beyond the scope of present-day computational power. 
However, there is a 3D model that can be solved exactly and that has a band gap in the infinite-size limit. 
The system is a metallic sphere in free space that is lossless at a particular frequency~\cite{Khurgin2010APL,Chew1987JCP}. 
A homogeneous medium with such a dielectric constant has a zero DOS at that frequency. 
In a sphere of finite radius $a$, however, the DOS will be nonvanishing and can be calculated by integrating the LDOS inside the sphere~\cite{Carminati2015SSR,LagTig96}.
%
\begin{eqnarray}
\rho_{\mathrm{sph}}(\omega) \equiv 
\frac{{\cal N}_{\mathrm{sph}}\left(\omega\right)}{V_{\mathrm{sph}}} = 
\frac{4\pi| \varepsilon|}{V_{\mathrm{sph}}}\int_{0}^{a}\rho_{\mathrm{sph}}\left(\omega, r\right) r^2 {\rm d } r,\label{eq:DOSSphere}
\end{eqnarray}
%
with $\varepsilon$ being the dielectric constant of the sphere,
${\cal N}_{\mathrm{sph}}\left(\omega\right)$ the total number of states per frequency inside the sphere, and $V_{\mathrm{sph}}\equiv 4\pi a^3/3$ the volume. 
%
\begin{figure}
	\begin{centering}
		\includegraphics[width=0.9\columnwidth]{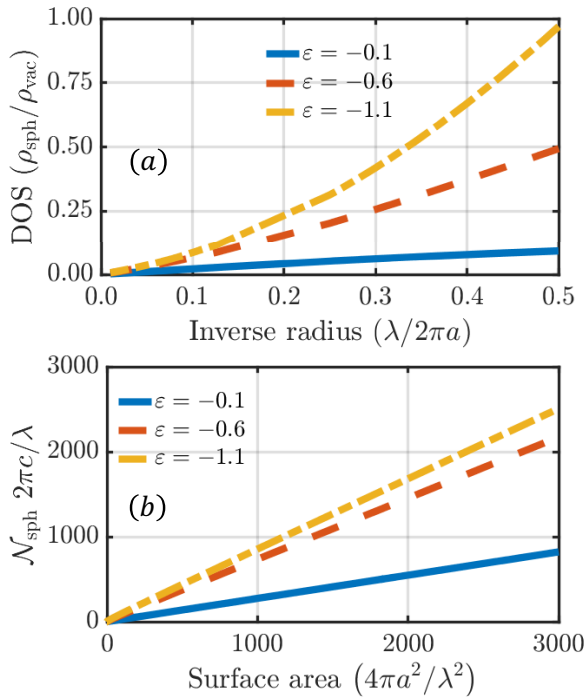}
		\par\end{centering}
	\caption{(a)~DOS, normalized to vacuum, inside a lossless metallic sphere with dielectric constant $\varepsilon = -0.1$ (blue), $\varepsilon = -0.6$ (red), and $\varepsilon = -1.1$ (yellow) as a function of the inverse size parameter $\lambda / 2 \pi a $, showing linear scaling for small values of inverse radius. 
The slopes of linear fits are $m=0.21$, $m=0.55$, and $m=0.64$ for $\varepsilon=-0.1$, $\varepsilon=-0.6$, and $\varepsilon=-1.1$, respectively.
(b)~Total number of states per frequency inside the sphere $\mathcal{N}_{\rm sph}$ as a function of the surface area. 
} 
\label{fig:DOSSphere}
\end{figure}
%
Figure~\ref{fig:DOSSphere}(a) shows the DOS inside the sphere $\rho_{\mathrm{sph}}(\omega)$ as a function of sphere radius at a fixed frequency for three typical negative values of the dielectric constant. 
In each case, the DOS inside the sphere decreases linearly with the inverse of the radius, which is the same behavior as the DOS found for 3D, 2D, and 1D photonic crystals~\cite{SI}. 
We can also calculate the total number of states ${\cal N}_{\rm sph}(\omega)$ for any radius, as shown in Fig.~\ref{fig:DOSSphere}(b). 
This figure, together with Eq.~(\ref{eq:DOSSphere}), demonstrates that the inverse linear scaling of the DOS is due to ${\cal N}\propto L^{D-1}$, which is the surface area of the sphere. 

Let us explore the origin of the universal $1/L$ finite-size scaling of the DOS. 
Light illuminating a photonic crystal with a frequency inside a stop gap will have its intensity exponentially attenuated with a characteristic length $L_\mathrm{B}$ called the Bragg length. We introduce a simple shell model where the contribution to the number of states \textit{inside} a finite-size crystal consists of a contribution of the bulk and a shell near the interface of depth $L_{\rm B}$. 
We assume that inside the bulk of the crystal the LDOS is constant: $\rho(\omega,{\bm r}) \approx \mathrm{constant} \equiv \rho_\mathrm{bulk} $. 
The spatial variation of the LDOS inside the shell is approximated by linear interpolation 
%
\begin{eqnarray}
\rho(z) &=& \rho_\mathrm{int} +  \frac{\rho_\mathrm{bulk} - \rho_\mathrm{int}}{L_\mathrm{B}}z, \ \ \ \mathrm{for}\ 0 < z \le   L_\mathrm{B},
\end{eqnarray}
%
where $\rho_\mathrm{int}$ is the LDOS at the vacuum-crystal interface and $z$ is defined as perpendicular to the interface that is located at $z=0$. 
Simple integration of the LDOS to obtain the DOS in the finite-support crystal and retaining the lowest power of $\LB/L$ results in~\cite{SI} 
%
\begin{eqnarray}
\rho & \approx & \left( 1 - \frac{f L_{\rm B}}{ 2L }\right) \rho_{\rm bulk} + 
 \frac{f  L_{\rm B}}{ 2L }  \rho_{\rm int}\label{eq:bulk_shell_dos},
\end{eqnarray}
%
where $f$ is the number of faces, with $f = 6$ in 3D, $f = 4$ in 2D, and $f = 2$ in 1D.
For a 1D photonic crystal with lattice spacing $d$, the Bragg length equals~\cite{NeveOz2004JAP,Lin1993OL,Vlasov1997PRB}
%
\begin{eqnarray}
 L_{\rm B} = \frac {2 d}{\pi}
 \frac{\omega_c}{\Delta \omega}\label{eq:bragg_length},
\end{eqnarray}
%
where $\omega_c$ is the central frequency of the gap and $\Delta \omega$ the width of the stopgap associated with the planes. 
For higher dimensions, the same equation pertains when for $d$ we take the smallest distance for a set of crystal planes.
In the case of a photonic band gap ($\rho_{\rm bulk} = 0$), the shell model leads to $1/L$ finite-size scaling, since $\rho \approx ( fL_{\rm B}/2L)\rho_{\rm int}$, confirming the scaling behavior found from the Lorentzian model.

With the shell model from Eq.~(\ref{eq:bulk_shell_dos}) ($\rho_\mathrm{int}\approx\rho_\mathrm{vac}$), we predict the slope in Fig.~\ref{fig:minDOS_IWP}(a) to be 1.54 
and that in Fig.~\ref{fig:minDOS_IWP}(b) to be 1.03.
These two 
slopes agree very well with the slopes 1.36 and 1.13, respectively, from our Lorentzian model [Eq.~(\ref{eq:pc_dos_definition_finite})]. 
The differences between the slopes predicted by the two models may be reduced 
by using more realistic values for the reflection coefficients
in the shell model. 
However, given its simplicity, we refrain from extending the shell model to include the complication of internal reflection.
We find that for all dimensions, both the Lorentzian and the shell model predict a $1/L$ scaling of the DOS. 
The fact that two such widely different methods lead to similar results is a strong evidence that they reveal the correct physical behavior.

We can also use the shell model Eq.~(\ref{eq:bulk_shell_dos}) to predict the scaling for the lossless metallic sphere, where we replace the Bragg length with the decay length $\lambda/( 4 \pi \sqrt{- \varepsilon})$ and use the exact value for $\rho_\mathrm{int}$~\cite{SI}. 
Our shell model predicts a $3\LB/a$ scaling of the DOS~\cite{SI}, leading to slopes in Fig.~\ref{fig:DOSSphere}(a) of 0.41, 0.68, and 0.72 for $\varepsilon = -0.1$, $\varepsilon = -0.6$, and $\varepsilon = -1.1$, respectively. 
These slopes agree well with the exact values 0.21, 0.55, and 0.64, demonstrating the accuracy of the shell model for the lossless sphere. 

We find that in finite-support photonic band gap crystals, almost all DOS contributions come from thin layers near the interfaces. 
It is well known that the LDOS in the center of a finite-support crystal scales exponentially with the size~\cite{Asatryan2001PRE,Ishizaki2009,Leistikow2011PRL}. 
However, for applications like the control of spontaneous emission of bulk emitters~\cite{Ogawa2004S,Leistikow2011PRL}, it is not the local but the global DOS that comes into play. 
The inverse linear scaling with size of this DOS indicates that realistic clusters have to be very large to show substantial photonic band gap effects.
Conversely, applications that aim to control spontaneous emission can benefit from a strongly modified DOS when light sources are removed from the thin layer near the crystal's surface.

%
\begin{acknowledgments}
This project is part of the program "Stirring of Light!" and the TTW Program 11985 that are part of the Nederlandse Organisatie voor Wetenschappelijk Onderzoek (NWO). 
A. P. M acknowledges a Vici grant from NWO. 
We thank Elahe Yeganegi for discussions and help with Fig.~\ref{fig:dos_cartoon}. 
\end{acknowledgments}

\end{document}